\documentstyle[12pt]{article}

\begin{document}

\title{Virtual Arbitrage Pricing Theory}
\author{
Kirill Ilinski
\thanks{E-mail: kni@th.ph.bham.ac.uk}
\\ [0.3cm]
{\small\it School of Physics and Space Research,
University of Birmingham,} \\
{\small\it Edgbaston B15 2TT, Birmingham, United Kingdom} 
}

\vspace{3cm}

\date{ }

\vspace{1cm}
 
\maketitle

\begin{abstract}
We generalize the Arbitrage Pricing Theory (APT) to include 
the contribution of virtual arbitrage opportunities. We model 
the arbitrage return by a stochastic process. The latter is incorporated 
in the APT framework to calculate the correction to the APT due to the virtual arbitrage
opportunities. The resulting relations reduce to the APT for an infinitely fast
market reaction or in the case where the virtual arbitrage is absent.
Corrections to the Capital Asset Pricing Model (CAPM) are also derived.
\end{abstract}

\section{Introduction}
Along with the Capital Asset Pricing Model (CAPM)~\cite{Sharp,Lintner,Mossin}
the Arbitrage Pricing Theory (APT)~\cite{Ross} is a standard 
model for determining 
the expected rate of return on individual stocks and on a portfolio
of stocks~\cite{portfolio}. 
In contrast to the CAPM, the APT does not rely on any assumptions
about utility functions or the assumption that agents consider only the mean
and variance of possible portfolios. Taking into account critical remarks about
the expected utility approach to the theory of the decision making under 
uncertainty, and the long running discussion about the definition of risk, this seems to 
be a considerable achievement. What the APT does imply are homogeneous expectations,
assumptions of linear factor weightings, a large enough number of
securities to eliminate the specific risk and {\it no-arbitrage} condition.
In simplified terms, the latter means that the return on a riskless portfolio
should be equal to the riskless rate of return which can be taken for a moment
equal to the rate of the return on a bank deposit (here, for sake of simplicity,
we assume perfect capital market conditions, i.e. single rate of lending and
borrowing without restrictions, absence of transaction costs and bid-ask 
spread; we touch on possible generalization in conclusion). 
Being a basis for the APT and derivative pricing~\cite{BS},
the no-arbitrage assumption appears to be very reasonable and robust.
Indeed, if any arbitrage possibility would exist then agents (arbitrageurs)
would use the opportunity to make an abnormal riskless profit which itself
will bring the system to the equilibrium and eliminate the arbitrage 
opportunity. Thus the arbitrage cannot exist for long and its lifetime
depends on a liquidity of the market. This, however, does not mean that
the arbitrage opportunities do not exist at all and cannot influence
the asset pricing violating the APT assumption. Moreover, there exist empirical
studies which point out the existence of virtual arbitrage opportunities
even for very liquid markets (order of 5 minutes for futures on
S\&P~\cite{Sofianos} and can be much longer for less liquid markets such as bonds
market and derivative market~\cite{der}). 
That is why in this paper we try
to overcome the no-arbitrage assumption and suggest a model to account for the
existence of virtual arbitrage opportunities and their influence on asset
pricing in the framework of the APT. To this end we follow the line 
introduced in Ref.~\cite{IS} to account the virtual arbitrage in 
derivative pricing.

One of the possible approaches to the problem of handling of virtual arbitrage
has been suggested in Ref~\cite{hep-th/9710148}
and based on a field-theoretical description of arbitrage opportunities 
and the corresponding money flows.
Being a consistent theory, the approach is however extremely
complicated and does not allow one yet to get simple analytical results which
would be easy understandable and handleable. That is why in this paper we
develop simplified tractable analytical version to account for virtual 
arbitrage opportunities which does not use complicated techniques and 
results in simple enough final formulas. 
 
The paper is organized as follows. In next section 
we derive an effective equation for a price of a riskless portfolio in
presence of virtual arbitrage opportunities.
There it is shown how the local 
arbitrage opportunities can be introduced in the model and how they change
the rate of return. In section 3 we derive equations of the APT under virtual
arbitrage.
These equations generalizes the APT relations and converge 
to them in the limit of absence of the arbitrage opportunities or 
infinitely fast market reaction. 
Section 4 is devoted to the corrections to the CAPM.
In conclusion we discuss the drawbacks 
of the model and possible ways to improve it.

\section{Effective equation for riskless portfolio in
presence of virtual arbitrage}

In this section we introduce a model for virtual arbitrage fluctuations and
describe their influence on return on riskless portfolio. Under no-arbitrage
assumption the rate of return on a riskless portfolio shall be equal to the
single riskless rate of return, which is equal to, for example, the return on
a bank deposit. However, if we allow existence of virtual arbitrage 
opportunities, the rate of return on a portfolio does not have to be equal
the rate of return on bank deposit, it is random (since
unpredictable) and might depend on the structure of a 
portfolio. Indeed, if, for example, the portfolio consists of only riskless 
assets (bank deposit or bond) there should be no arbitrage opportunities with 
this portfolio and the rate of return on this portfolio, by definition, shall be 
equal to the rate of return on a bank deposit $r_0$. In the general case of 
a riskless portfolio which consists of diversified risky assets the existence
of the arbitrage fluctuations would depend on the portfolio composition. 
To give an intuitive example, we can imagine that 
if some market sector is more attractive for investors,
the trades and resulting arbitrage fluctuations will occur more often.
This will result in more arbitrage opportunities for portfolios containing 
the assets of the sector and the corresponding composition dependence.

Let us consider riskless portfolio $\Pi$ which is created by $N+1$ assets
with fractions $\{x_i\}_{i=0}^{N}$. In the case of no-arbitrage the portfolio
$\Pi$ would satisfy the following equation:
\begin{equation}
\frac{d \Pi}{dt} - r_0 \Pi = 0 \ , \qquad \Pi (0) = 1 
\label{no-arbitrage}
\end{equation}
where $r_0$ is riskless interest rate on bank deposit.
However, in case of the virtual arbitrage RHS of Eq(\ref{no-arbitrage})
shall be changed to
$$
{\cal R}(t,\Pi) \Pi
$$
where ${\cal R}(t,\Pi)$ represents the virtual arbitrage return.
To find an expression for ${\cal R}(t,\Pi)$
let us imagine that at some moment of time $\tau < t$
a fluctuation of the return (an arbitrage opportunity) appeared in the market. 
We then denote this instantaneous arbitrage return 
as $\nu (\tau,\Pi)$.
Arbitragers would react to this circumstance and act in such a way that the 
arbitrage gradually disappears and the market returns to its equilibrium state,
i.e. the absence of the arbitrage.
For small enough fluctuations it is natural to assume that the arbitrage return
${\cal R}$ (in absence of other fluctuations) evolves according to the following
equation:
\begin{equation}
\frac{{\rm d}{\cal R}}{{\rm d}t} = - \lambda {\cal R} \ , \qquad 
{\cal R}(\tau) = \nu (\tau,\Pi)
\label{R1}
\end{equation}
with some parameter $\lambda$ which is characteristic for the market.
This parameter can be either estimated from a microscopic theory like
\cite{hep-th/9710148} or can be found from the market using
an analogue of the fluctuation-dissipation theorem~\cite{Landau}. 
In the last case the parameter $\lambda$ can be estimated from the market data 
as
\begin{equation}
\lambda = 
-\frac{1}{(t - t^{\prime})}\log 
\left[
\left\langle 
\left( r_{\Pi} - r_0 \right) (t)
\left( r_{\Pi} - r_0 \right) (t^{\prime})
\right\rangle_{\rm market} /  
\left\langle
\left(
r_{\Pi} - r_0
\right)^{2}(t)
\right\rangle_{\rm market}
\right] \quad , t > t^{\prime}
\label{lambda}
\end{equation}
and may well be a function of time,
portfolio composition and even prices of assets. 
In what follows we however consider $\lambda$ as a constant to
get simple analytical formulas for APT corrections. 
The generalization to the case of time-dependent parameters 
is straightforward.

The solution 
of Eqn(\ref{R1}) gives us ${\cal R}(t,\Pi) = \nu(\tau,\Pi) 
{\rm e}^{-\lambda (t-\tau)}$ which,
after summing over all possible fluctuations, 
leads us to the following expression for the arbitrage return
\begin{equation}
{\cal R}(t,\Pi) \equiv \int_{-\infty}^{t} e^{-\lambda (t-\tau)} 
\nu (\tau,\Pi) d\tau \ .
\label{p2}
\end{equation}
To specify the stochastic 
process $\nu(t,\Pi)$ we assume that the fluctuations at different times 
are independent and form the white noise with a variance 
$\Sigma^2 (\Pi)$ which depends on the structure of the portfolio $\Pi$:
\begin{equation}
\langle\nu(t,\Pi)\rangle_{\nu} =0 \quad \ , \qquad 
\langle\nu(t,\Pi) \nu(t^{\prime}, \Pi)\rangle_{\nu} = 
\Sigma^2(\Pi) \cdot \delta (t-t^{\prime}) \ ,
\label{nu}
\end{equation}
where, again, the parameter $\Sigma^2(\Pi)$ can be taken from the market:
\begin{equation}
\Sigma^2(\Pi)/2 \lambda =   
\left\langle\left(
r_{\Pi} - r_0
\right)^{2} \right\rangle_{\rm market} \ .
\label{sigma}
\end{equation}
To simplify the consideration we assume here that the quantity does not 
depend on time but this limitation may be straightforwardly overcome 
(see discussion in the conclusion).

Since we introduced the stochastic arbitrage return  ${\cal R}(t,\Pi)$,
Eqn(\ref{no-arbitrage}) has to be substituted with the following equation:
\begin{equation}
\frac{d \Pi}{dt} - r_0 \Pi = {\cal R}(t,\Pi) \Pi
\label{arbitrage1}
\end{equation}
or, in the integral form,
\begin{equation}
\Pi = 
\int_{-\infty}^{t} G(t,t^{\prime}) {\cal R}(t^{\prime},\Pi) \Pi (t^{\prime}) 
d t^{\prime}
\label{arbitrage2}
\end{equation}
where $G(t,t^{\prime}) = \Theta (t-t^{\prime}) e^{r_0 (t-t^{\prime})}$
is Green function of the problem:
$$
(\frac{d }{dt} - r_0) G(t,t^{\prime}) = \delta (t-t^{\prime}) \ ,
\qquad G(t,t^{\prime}) |_{t<t^{\prime}} = 0 \ .
$$
We can iterate Eq(\ref{arbitrage1}) and substitute Eq(\ref{arbitrage2})
in RHS of Eq(\ref{arbitrage1}) which gives
\begin{equation}
\frac{d \Pi}{dt} - r_0 \Pi = {\cal R}(t,\Pi) 
\int_{-\infty}^{t} G(t,t^{\prime}) {\cal R}(t^{\prime},\Pi) \Pi (t^{\prime}) 
d t^{\prime} \ .
\label{arbitrage3}
\end{equation}

Our next step is to average Eq (\ref{arbitrage3}) over virtual arbitrage fluctuations
and to obtain the effective pricing equation for the average price $\bar{\Pi}$.
At the first order in $1/\lambda$ it can be written as
\begin{equation}
\frac{d \bar{\Pi}}{dt} - r_0 \bar{\Pi} =  
\int_{-\infty}^{t} G(t,t^{\prime}) K(t,t^{\prime},\bar{\Pi}) \bar{\Pi} 
(t^{\prime}) d t^{\prime} \ .
\label{eff1}
\end{equation}
where the kernel 
$$
K(t,t^{\prime},\bar{\Pi}) =
\langle {\cal R}(t,\Pi) {\cal R}(t^{\prime},\Pi)\rangle_{\nu}
$$ 
is given by the expression~\cite{IS}:
\begin{equation}
K(t,t^{\prime}) =
\frac{\Sigma^2 (\bar{\Pi})}{2\lambda}
\theta(t-t^{\prime})
{\rm e}^{-\lambda(t-t^{\prime})}
 + \frac{\Sigma^2 (\bar{\Pi})}{2\lambda}\theta(t^{\prime}-t) 
{\rm e}^{-\lambda(t^{\prime}-t)}
\label{K}
\end{equation}
which can be easily obtained from Eq(\ref{nu}). 
Collecting everything together, Eqs(\ref{eff1},\ref{K}) are used in place of 
Eq(\ref{no-arbitrage}) in situations where virtual arbitrage opportunities 
exist.

To solve Eq(\ref{eff1}) we first notice that its RHS can be written as
$$
\int_{-\infty}^{t} G(t,t^{\prime}) K(t,t^{\prime},\bar{\Pi}) \bar{\Pi} 
(t^{\prime}) d t^{\prime} = 
\int_{-\infty}^{t} e^{-\lambda (t-t^{\prime})} 
e^{r_0 (t- t^{\prime})} \frac{\Sigma^{2}(\bar{\Pi})}{2\lambda} 
\bar{\Pi} (t^{\prime}) d t^{\prime} \  .
$$
We now make an approximation $\bar{\Pi} (t^{\prime}) =
e^{r_0 t^{\prime}}$ because the correction to this expression is order of
$\frac{\Sigma^{2}(\bar{\Pi})}{2\lambda}$ and is irrelevant for our consideration
of RHS of Eq(\ref{eff1}) since it was derived within this order of accuracy.
Such approximation results in the following relation:
$$
\int_{-\infty}^{t} G(t,t^{\prime}) K(t,t^{\prime},\bar{\Pi}) \bar{\Pi} 
(t^{\prime}) d t^{\prime} = 
\frac{\Sigma^{2}(\bar{\Pi})}{2\lambda ^2} e^{r_0 t} \ .
$$
If we substitute it into Eq (\ref{eff1}) this gives us the following
approximate differential equation for the average portfolio price:
\begin{equation}
\frac{d \bar{\Pi}}{dt} - r_0 \bar{\Pi} = 
\frac{\Sigma^{2}(\bar{\Pi})}{2\lambda^2} e^{r_0 t}
\label{eff2}
\end{equation}
which has the solution:
$$
\bar{\Pi} (t) = (1+ \frac{\Sigma^{2}(\bar{\Pi})}{2\lambda ^2} t) e^{r_0 t}
$$
or, to the same level of accuracy,
\begin{equation}
\bar{\Pi} (t) = 
e^{(r_0 + \frac{\Sigma^{2}(\bar{\Pi})}{2\lambda ^2} ) t} \ .
\label{Pi}
\end{equation}
The latter expression we use in the next section 
in place of the 
no-arbitrage expression ${\Pi} (t) = e^{r_0  t}$ in the Arbitrage 
Pricing Theory.

Let us emphasize that we, as well as an investor, 
are interested in the average value of the portfolio
and the corresponding rate of return on the average portfolio rather than in 
average rate of return on the riskless portfolio which is exactly equal to
$r_0$ and, in fact, is not influenced by the arbitrage. 
Indeed, the only valuable and material thing for an investor 
is the average value of her investment. She is not interested in a mathematical
quantity which is not actually connected with her wealth.
The difference between the two returns is the first effect which is
solely due to the presence of the virtual arbitrage.

\section{Corrections to APT}
Let us first remind ourselves of the standard derivation of Arbitrage Pricing 
Theory relations. The first assumption below APT is the existence of $M$ 
fluctuating leading parameters, or factors,
$\{\xi_{j}\}_{j=1}^{M}$ which define the prices
of all $N+1$ assets such that
$$
\langle \xi_{j} \rangle_{\xi} = 0 \ , \qquad
\langle \xi_{i} \xi_{j} \rangle_{\xi} = \delta_{ij}
$$
and
\begin{equation}
r_i = 
\bar{r}_i + \sum_{j} b_{ij} \xi_j + \epsilon_i 
\label{Pi1}
\end{equation}
with $b_{ij}=\langle r_i-\bar{r}_i,\xi_j\rangle_{\xi}$ as a measure 
of influence of $j$-th parameter on $i$-th asset and $\epsilon_i$ is 
a residual risk which is independent on the factors and can
be eliminated in large portfolios.
To form a riskless portfolio we pick up such fractions $\{x_i\}$ that
$$
\sum_{i} x_i b_{ij} = 0 
$$
for any $j=1,...,M$ and they obey normalization
condition $\sum_{i} x_i = 1$.
We assume that $0$-th asset is riskless,
e.g. bank deposit or treasury bonds.

The return on the portfolio is $\sum_{i} r_i x_i = \sum_{i} \bar{r}_i x_i$
and {\it has to be equal} to the risk-free interest rate $r_0$ in the 
case of {\it no-arbitrage}. This leads to the equation
\begin{equation}
\sum_{i} \bar{r}_i x_i = r_0
\label{ap}
\end{equation}
or, after a change of variables $x_i = y_i + \delta_{0i}$,
$$
\sum_{i} \bar{r}_i y_i = 0
$$
subject to the constraints
$$
\sum_{i} y_i b_{ij} = 0 \ \ \forall j \ , \quad \sum_{i} y_i = 0 \ .
$$
The solution of the equation for $\bar{r}_i$ then can be written as
\begin{equation} 
\bar{r}_i = \alpha + \sum_{j} b_{ij} \gamma_j \ .
\label{APT}
\end{equation}
where $\alpha$ and $\{\gamma_j\}$ are set of independent parameters.
Since the $0$-th asset was chosen as risk-free, the coefficients
$b_{0j}=0$ for all $j$ and $\alpha =r_0$. Eq(\ref{APT}) constitutes the
Arbitrage Pricing Theory relations.

To generalize the APT relations to the case of virtual arbitrage 
we have to deal with average portfolios considered in the previous
section. We define the rates of the return 
$\{\bar{r}^{\prime}_k\}_{k=0}^{N}$ on $k$-th component of 
an average riskless portfolio as
$$
\bar{\Pi}(t) = e^{\sum_{i} \bar{r}^{\prime}_i x_i t} \ .
$$
These are the quantities which have to substitute the rates of return in 
the APT relations.

To find an expression for $\{\bar{r}^{\prime}_k\}_{k=0}^{N}$ we have to
substitute $r_0$ in Eq(\ref{ap}) on the return to the riskless portfolio
derived under virtual arbitrage consideration in the previous section, i.e.
on $(r_0 + \frac{\Sigma^{2}(\bar{\Pi})}{2\lambda ^2} )$ as follows 
from Eq(\ref{Pi}). This gives us the equation for $\bar{r}^{\prime}_i$:
\begin{equation}
\sum_{i} \bar{r}^{\prime}_i y_i = \frac{\Sigma^{2}(\bar{\Pi})}{2\lambda ^2}
\label{price}
\end{equation}
subject to constraints
\begin{equation}
\sum_{i} y_i b_{ij} = 0 \ \ \forall j \ ,  \quad \sum_{i} y_i = 0 \ .
\label{constr}
\end{equation}
To simplify the following consideration we introduce a convenient basis
in the portfolio space such that the above constraints will take the form
$$
\eta_i = 0 \ , \ \  0 \leq i \leq M\ 
$$
in this new basis while $\{\eta_j\}_{j=M+1}^{N}$ will be coordinates in the 
subspace of risk-free portfolios. To this end we first introduce vectors
$\{{\bf e}^{\prime}_i\}_{i=0}^{M}$:
$$
e_{0,i}^{\prime}=1\ \forall i \ , \ e_{i,j}^{\prime}=b_{ij}\  \forall i \ ,
1\leq j \leq M \ .
$$
These $M+1$ vectors are linear independent (otherwise it would be linear 
dependence for the factors and $M+1$ constraints (\ref{constr})
would not be independent). The rest of $N-M$ linear independent  vectors 
$\{{\bf e}^{\prime}_i\}_{i=M+1}^{N}$ can be chosen arbitrary but linear 
independent with $\{{\bf e}^{\prime}_i\}_{i=0}^{M}$. For example, if
$$
b_{ij}\neq 0 \ , \ \ \ \forall j \ \mbox{and } \ 0<i < M+1  
$$
then one possible choice could be
$$
e_{i,j}^{\prime}=\delta_{ij}\ \ \ \ \forall j \ ,
M+1\leq i \leq N+1 \ .
$$
The next step is the Gram-Schmidt orthogonalisation and normalisation 
of the vector set $\{{\bf e}^{\prime}_i\}_{i=0}^{N}$ which gives us the 
basis set $\{{\bf e}_i\}_{i=M+1}^{N}$. We shall start the orthogonalisation
with the vector ${\bf e}_{0}^{\prime}$:
$$
{\bf e}_{0}= \frac{{\bf e}_{0}^{\prime}}{ 
\sqrt{({\bf e}_{0}^{\prime},{\bf e}_{0}^{\prime})}} \ ,
$$
proceed with the vector ${\bf e}_{1}^{\prime}$ as
$$
{\bf e}_{1} = \frac{{\bf e}_{1}^{\prime} - 
({\bf e}_{1}^{\prime},{\bf e}_{0}){\bf e}_{0}}{
\sqrt{({\bf e}_{1}^{\prime} - 
({\bf e}_{1}^{\prime},{\bf e}_{0}){\bf e}_{0},
{\bf e}_{1}^{\prime} - 
({\bf e}_{1}^{\prime},{\bf e}_{0}){\bf e}_{0})}
}
$$
and carry on following the standard procedure:
$$
{\bf e}_{j} = \left({\bf e}_{j}^{\prime} - \sum_{k=0}^{j-1} 
({\bf e}^{\prime}_{j},{\bf e}_{k}) {\bf e}_{k}\right) / 
\left( 
({\bf e}_{j}^{\prime},{\bf e}_{j}^{\prime}) - 
\sum_{k=0}^{j-1} ({\bf e}_{j}^{\prime},{\bf e}_{k})^2
\right)^{1/2} \ .
$$
It is easy to check now that in this new basis 
$\{{\bf e}_i\}_{i=0}^{N}$ the constraints (\ref{constr}) can be represented
as a zero projection along the first $M+1$ basis vectors, i.e. as
$$
\eta_i = 0 \ ,  \ \ \  0 \leq i \leq M\ 
$$
where $\eta_i$ are coordinates in the new basis. Indeed, for any vector 
${\bf y}$ the conditions
$$
({\bf y},{\bf e}^{\prime}_{j}) = 0 \ , \ \  0\leq j \leq M
$$
are exactly equivalent to the constraints (\ref{constr}). This remains true 
also after the orthogonalisation 
$$
({\bf y},{\bf e}_{j}) = 0 \ , \ \  0\leq j \leq M
$$
due to step-by-step nature of the procedure. Therefore any vector
$$
{\bf y} = \sum_{j=M+1}^{N+1} \eta_j {\bf e}_{j}
$$
represents a risk-free portfolio and the set $\{{\bf e}_j\}_{j=M+1}^{N}$ is
a basis in the subspace of riskless portfolios.

To express the riskless portfolios basis  in terms of the original assets we
introduce the rotation matrix $U=||U_{ik}||$:
\begin{equation}
U = \left( 
\begin{array}{c}
{\bf e}_{0} \\
{\bf e}_{1} \\
... \\
{\bf e}_{N}
\end{array}
\right) \ .
\label{U}
\end{equation}
Using this matrix the portfolio represented by the
basis vector ${\bf e}_i$ can be constructed by
collecting $j$-th assets ($j=0,..,N+1$) with the fractions 
$e_{i,j}+\delta_{0j}$. 

Now we are ready to discuss the possible form of
$\Sigma^{2}(\bar{\Pi})$ as a function of portfolio structure, i.e.
as a function of $\{y_i\}_{i=0}^{N}$ or $\{\eta_i\}_{i=M+1}^{N}$. 
First of all we will not
allow the existence of arbitrage fluctuations for a portfolio which consists
of pure bonds, i.e. when all $\eta_i$ equal to zero. This provides that the
rate of return on the bank deposit is actually equal to $r_0$, does not
have any arbitrage corrections and our use of $r_0$ is self-consistent.
It means that series expansion of $\Sigma^{2}(\eta)$ cannot contain a 
constant term. Second, since the virtual arbitrage return is brought 
in by  portfolio $\eta$, the arbitrage return on the  portfolio $-\eta$
has to be equal to minus the latter and $\Sigma^2(\eta)=\Sigma^2(-\eta)$. 
This results in the absence of all odd terms. Therefore the
series expansion of $\Sigma^{2}(\bar{\Pi})$ has the form:
$$
\Sigma^{2}(\bar{\Pi}) = 
\sum_{i,j=M+1}^{N} \Sigma^{2}_{ij} \eta_i \eta_j  + ... \quad ,
$$
or 
$$
\Sigma^{2}(\bar{\Pi}) = 
\sum_{i,j=M+1}^{N} \Sigma^{2}_{ij} \eta_i \eta_j  + ... \quad ,
\qquad \Sigma^{2}_{i...k} = 0 \ \ \mbox{for } \ \ 0 \leq i,...,k \leq M \ .
$$
In what follows we keep only first nontrivial term, 
$({\bf \eta },\Sigma^2{\bf \eta })$,
$\Sigma^2\equiv ||\Sigma^2_{ik}||$ in the expansion though
the final result will be valid for the general case too.

Returning back to pricing equation (\ref{price}) we substitute 
${\bf y} = U{\bf \eta}$,  
$\Sigma^{2}(\bar{\Pi}) = ({\bf \eta },{\Sigma^{2}}{\bf \eta })$,
$\sum_{i} \bar{r}_i y_i = ({\bf \bar{r}}, {\bf y})$
we come to the pricing equation for ${\bf \bar{r}}$:
$$
({\bf \bar{r}}, U{\bf \eta}) = ({\bf \eta },{ \Sigma^{2}}{\bf \eta })
$$
subject to constraints
$$
\eta_i = 0  \ , \ \ \   0 \leq j \leq M \ .
$$
The solution of the problem can be easily found as
\begin{equation} 
\bar{r}_i = \alpha + \sum_{j} b_{ij} \gamma_j + 
\sum^{N}_{i,k,l=0} U_{ik} \frac{\Sigma^{2}_{kl}}{2\lambda ^2} \eta_l\ .
\label{VAPT0}
\end{equation}

At this point we face an unpleasant problem: a return on an asset in the 
portfolio depends on the structure of the portfolio. This is the second
effect which stems from the virtual arbitrage presence. However unpleasant it is,
it is hardly surprising since it was derived under the assumption of virtual 
arbitrage which is portfolio-dependent. Intuitively it is also clear: if an 
asset constitutes a part of a ``hot" arbitrage portfolio its rate of return 
will be different from assets in ``quiet" portfolios {\it ceteris paribus}.

To find an average growth factor for the $i$-th asset we have to take a 
sum of $e^{\bar{r(\Pi)}^{\prime}_i}$ weighted with probabilities of appearance of 
the portfolio $\Pi$, i.e. to calculate the average value
$$
\langle e^{\bar{r}^{\prime}_i(\Pi)}\rangle_{\Pi} \equiv
e^{\bar{r}_i}
$$
which defines the rate of growth $\bar{r}_i$ of the average growth 
factor. This $\bar{r}_i$ is a counterpart of the average rate of return
under the virtual arbitrage presence since it characterizes how fast 
$i$-th asset grows in average. It is easy to see that in the first 
nontrivial order 
$$
e^{\bar{r}_i} = e^{\alpha + \sum_{j} b_{ij} \gamma_j} \left( 1+
\sum^{N}_{k,m,l,l^{\prime}=0} 
U_{ik} U_{im} \frac{\Sigma^{2}_{kl} \Sigma^{2}_{ml^{\prime}}}{8\lambda^4} 
\langle \eta_{l} \eta_{l^{\prime}} \rangle_{\Pi}
\right)
$$
and, hence, with the same measure of accuracy the rate of growth is given by
\begin{equation}
\bar{r}_i =
\alpha + \sum_{j} b_{ij} \gamma_j +
\sum^{N}_{k,m,l,l^{\prime}=0} 
U_{ik} U_{im} \frac{\Sigma^{2}_{kl} \Sigma^{2}_{ml^{\prime}}}{8\lambda^4} 
\langle \eta_{l} \eta_{l^{\prime}} \rangle_{\Pi}
\label{VAPT}
\end{equation}
Since the $0$-th asset was chosen as risk-free, the coefficients
$\beta_{0j}=0$ for all $j$ and 
\begin{equation} 
\bar{r}_i = r_0 + \sum_{j} b_{ij} \gamma_j + 
\sum^{N}_{k,m,l,l^{\prime}=0} 
U_{ik} U_{im} \frac{\Sigma^{2}_{kl} \Sigma^{2}_{ml^{\prime}}}{8\lambda^4} 
\langle \eta_{l} \eta_{l^{\prime}} \rangle_{\Pi}  -
\sum^{N}_{k,m,l,l^{\prime}=0} 
U_{0k} U_{0m} \frac{\Sigma^{2}_{kl} \Sigma^{2}_{ml^{\prime}}}{8\lambda^4} 
\langle \eta_{l} \eta_{l^{\prime}} \rangle_{\Pi} \ .
\label{VAPT1}
\end{equation}
The last equation looks quite complicated. It
can be presented in a more compact form using matrix notations and
the following matrix $\Delta$:
$$
\Delta_{ik} = \frac{1}{4\lambda^2}\left\langle 
\frac{\partial {\Sigma^{2}(\eta)}}{\partial \eta_i} 
\frac{\partial {\Sigma^{2}(\eta)}}{\partial \eta_k}
\right\rangle_{\Pi} \ .
$$
The pricing relations then can be rewritten as
\begin{equation} 
\bar{r}_i = r_0 + \sum_{j} b_{ij} \gamma_j + 
\frac{1}{2\lambda^2} (U\cdot \Delta \cdot U^{-1})_{ii} -
\frac{1}{2\lambda^2} (U\cdot \Delta \cdot U^{-1})_{00} \ .
\label{VAPT2}
\end{equation}
Eqns (\ref{VAPT1},\ref{VAPT2}) gives a generalization of the 
APT relations in the case where virtual arbitrage opportunities exist. 
It is easy to see that the
corrections are proportional to the square of the product of the variance of 
the arbitrage fluctuations and the square of a characteristic lifetime of the
arbitrage. This results in the disappearance of the correlations in the case 
of infinitely fast market reaction ($\lambda \rightarrow \infty$) or an
absence of the arbitrage ($\Sigma \rightarrow 0$) and the reduction of 
Eqns(\ref{VAPT1},\ref{VAPT2}) to the APT expression (\ref{APT}).

\section{Simplest application: correction to CAPM}

In this short section we concentrate on
a particular case of APT with the only one
leading parameter which will be chosen to be a random part of a 
return on a market portfolio. In this case the standard APT relations 
are reduced to the CAPM equations.

In the case of only one leading parameter chosen to be 
a random part of a return on the market portfolio
$\xi\equiv (r_{m}-\bar{r}_m)/\sigma_{m}$,
Eqn (\ref{VAPT2}) can be rewritten as
\begin{equation} 
\bar{r}_i = 
r_0 + b_{i} \gamma
\frac{1}{2\lambda^2} (U\cdot \Delta \cdot U^{-1})_{ii} -
\frac{1}{2\lambda^2} (U\cdot \Delta \cdot U^{-1})_{00}
\label{CAPM1}
\end{equation}
where $b_i = \langle r_i-\bar{r}_i,r_m-\bar{r}_m \rangle/\sigma_m$. 
The next step is to
introduce the market portfolio fractions $\theta_i$ and find an expression
for the average rate of return on the market portfolio 
$\sum_i \theta_i \bar{r}_i$ using Eqn (\ref{CAPM1}):
$$
\bar{r}_m = r_0 + \sum_i \theta_i b_i \gamma +
\frac{1}{2\lambda^2} u^m -
\frac{1}{2\lambda^2} (U\cdot \Delta \cdot U^{-1})_{00}
$$
where we introduce a notation 
$u^m = \sum_i (U\cdot \Delta \cdot U^{-1})_{ii} \theta_i$.
This allow us to define a value of the variable $\gamma$:
$$
\gamma = \frac{1}{\sigma_m} \left(
\bar{r}_m - {r}_0 + \frac{1}{2\lambda^2} (U\cdot \Delta \cdot U^{-1})_{00} -
\frac{1}{2\lambda^2} u^m
\right) \ ,
$$
and to find a final expression for the CAPM generalized to the virtual 
arbitrage assumption:
\begin{equation} 
\bar{r}_i = r_0 + \beta_{i} \left(
\bar{r}_m - {r}_0\right) + 
\frac{\beta_{i}}{2\lambda^2} \left(
(U\cdot \Delta \cdot U^{-1})_{00} - u^m
\right) +
\frac{(U\cdot \Delta \cdot U^{-1})_{ii}}{2\lambda^2}  -
\frac{(U\cdot \Delta \cdot U^{-1})_{00}}{2\lambda^2} \ 
\label{CAPM2}
\end{equation}
with the standard definition 
$\beta_i = \langle r_i - \bar{r}_i, r_m - \bar{r}_m \rangle /
\sigma_m^2$. The first two terms represent the CAPM under the 
no-arbitrage assumption
and the remaining terms are the virtual arbitrage corrections. Everything
said in the last paragraph of the previous section can be repeated with 
respect to Eqn(\ref{CAPM2}) for the virtual arbitrage generalization of 
the CAPM.

\section{Discussion}
In conclusion, we introduced a virtual arbitrage opportunities in the 
framework of the Arbitrage Pricing and derived corrections to the APT 
pricing relations. These corrections disappear as $\Sigma \rightarrow 0$ or 
$\lambda \rightarrow 0$ i.e. when there is no arbitrage fluctuations or 
the speed of market reaction to the mispricing is infinite (which corresponds 
to extremaly liquid market). 

In the course of the analysis we faced two major effects which are specific
for the presence of the arbitrage. The first one is a necessity to analyze
growth factors rather than the the corresponding rates of return. The
second effect is a dependence of a growth factor of an asset on the structure
of a riskless portfolio which includes the asset. This forced us to introduce
an average over all riskless portfolios and reduced the problem to a 
calculation of the matrix $\Delta$
\begin{equation}
\Delta_{ik} = \frac{1}{4\lambda^2} \left\langle 
\frac{\partial \Sigma^{2}(\eta)}{\partial \eta_i}  \ 
\frac{\partial \Sigma^{2}(\eta)}{\partial \eta_k}
\right\rangle_{\Pi} 
\label{delta}
\end{equation}
which is essentially a new element. We already mentioned that $\Sigma^2(\Pi)$
can be obtained from the market using Eqn(\ref{sigma}). It means that $\Delta$,
in principle, can also be obtained from the market following (\ref{delta}). 
However the most economical procedure of the evaluation still has to be 
worked out. It might, for example, be reasonable to consider some most frequent
arbitrages and reduce the space of the relevant riskless portfolios. Another 
option is to study thhe main factors generating arbitrage opportunities and carry
out a factor analysis similar to the original APT. At the moment we leave this
as an open question which requires further investigation.
Now let us discuss some obvious drawbacks of the model and ways to improve it. 

First of all, we have to admit immediately that it would be more difficult to
carry out empirical tests of the Virtual Arbitrage Pricing Theory than the APT
because the additional terms in Eqn(\ref{VAPT1}) do not make life 
easier in any respect while even empirics for the CAPM and the
APT~\cite{emp} have not produced yet any decisive result.
Furthermore, all critical comments of APT analysis 
can be forwarded to the presented model, except for the 
no-arbitrage constraint. Homogeneous expectations, 
absence of clear recipe of how to measure the factors $\xi_i$ and
possible time dependence of the parameters of the model are three obvious
points to criticize. These are not new problems and many efforts 
to overcome these difficulties have been undertaken. 
It is possible to demonstrate that the virtual arbitrage 
model can be improved in the same manner by these methods as they succeed
for no-arbitrage APT analysis. This, in principle, allows one to include 
the transaction costs, taxes and market imperfection in
the present model by rederiving the "bare" APT relations and then to add the
virtual arbitrage in the consideration with minor modifications. 

The second point concerns the market reaction to the arbitrage opportunity, or,
qualitatively the form of  ${\cal R}(t,\Pi)$ in Eqn(\ref{p2}). 
It may be argued that the market reaction is not exponential as assumed 
in Eqn(\ref{R1}), but has another functional dependence. This dependence 
can be found from statistical analysis
of the stock prices and then included in the equation for
${\cal R}(t,\Pi)$ (in particular, the functional dependence can 
change with time, for example $\lambda$ in (\ref{R1}) can be a function of 
$\tau$). It certainly complicates the model but leaves the general framework
intact.

Another point to consider is the absence of correlations between virtual
arbitrage opportunities which we assumed in the text, i.e. the white noise
character of the process $\nu(t,\Pi)$. It is clear that some correlations
can be easily included in the model by a proper 
substitution  of the relations in Eqn(\ref{nu}). 
Such generalizations, though making
the analytical study almost impossible, allow one to proceed with numerical 
analysis for the model.

Finally, the model contains new parameters such as $\Sigma$ and $\lambda$
defined by Eqns (\ref{sigma}) and (\ref{lambda}) from the market data.
It might appear that time-independent parameters
are not good enough approximation and both $\Sigma$ and $\lambda$ are functions 
of time as a consequence of market intermittent busts of activity. Furthermore,
the parameter $\lambda$ might actually depends on a structure of the portfolio
exactly in the same manner as $\Sigma$ does because of particular sector
preferences of arbitrageurs. Both situations can be processed in a 
similar manner to the simplest case we considered above but it will 
make the final results much more complicated.

\end{document}